\documentclass[letterpaper,twocolumn,10pt]{article}
\usepackage{usenix-2020-09}

\usepackage{tikz}
\usetikzlibrary{patterns}
\usepackage{alltt}
\usepackage{amsmath,amssymb}
\usepackage{balance}
\usepackage{booktabs}
\usepackage[capitalise,noabbrev]{cleveref}
\usepackage{pifont}
\usepackage{subcaption}
\usepackage{xspace}
\usepackage[normalem]{ulem}
\usepackage{listings}
\usepackage{totpages}
\usepackage{tabularx}
\usepackage{tikz-network}
\usepackage[nounderscore]{syntax}
\usepackage{colortbl}
\usepackage[framemethod=tikz]{mdframed}

\usepackage{listings}
\usepackage{solarized}      

\lstdefinelanguage{P4}
{
  numberstyle=\tiny,
  showspaces=false,
  showtabs=false,
  tabsize=2,
  columns=flexible,
  keepspaces=true,
  language={Java},
  numbers=left,
  xleftmargin=0pt,
  basicstyle=\ttfamily\footnotesize,
  numberstyle=\scriptsize\color{gray},
  showstringspaces=false,
  upquote=true,
  xleftmargin=1.2em,
  framexleftmargin=1.5em,
  morekeywords={table, reads, if, elif, else, for, int, actions, and, max_size, default_action, alts, control, apply, action, reaction, malleable, value, field, fields, metadata, header_type, width, uint16_t, init, ing, egr, reg, field_list, drop, field_list_calculation, parser, extract, return, default, select, exact, lpm},
  morecomment=[l]{//},
  morecomment=[s]{/*}{*/},
  morestring=[b]",
  moredelim=*[is][\textcolor{scyan}]{\%}{\%},
  moredelim=*[is][\textcolor{sblue}]{\#}{\#},
  moredelim=*[is][\textcolor{sorange}]{|}{|},
  moredelim=*[is][\textcolor{sbase00}]{?}{?},
  moredelim=**[is][\color{sbase3light}]{<>}{<>}
}

\lstdefinelanguage{python}
{
  morekeywords={if, elif, while, not, else, for, int, in, continue, uint16_t},
  morecomment=[l]{//},
  morestring=[b]",
  moredelim=*[is][\textcolor{scyan}]{\%}{\%},
  moredelim=*[is][\textcolor{sblue}]{\#}{\#},
  moredelim=*[is][\textcolor{sorange}]{|}{|},
  moredelim=*[is][\textcolor{smagenta}]{;}{;},
  moredelim=*[is][\textcolor{sviolet}]{?}{?},
  moredelim=**[is][\color{sbase3light}]{<>}{<>}
}

\newcommand{\topheading}[1]{\noindent\textbf{#1.}}
\newcommand{\heading}[1]{\vspace{6pt}\noindent\textbf{#1.}}
\newcommand{\headingraw}[1]{\vspace{6pt}\noindent\textbf{#1}}

\newcommand{\noheading}[0]{\vspace{6pt}\noindent}

\newcommand{\code}[1]{{\small\texttt{#1}}}
\newcommand{\naive}[0]{na\"ive\xspace}
\newcommand{\Naive}[0]{Na\"ive\xspace}

\newcommand{\name}[0]{\textit{FP4}\xspace}

\newcolumntype{Y}{>{\centering\arraybackslash}X}

\mdfdefinestyle{background}{backgroundcolor=lightorange,innerrightmargin=0cm,innertopmargin=-0.1cm,innerbottommargin=-0.10cm,leftmargin=+0cm,innerleftmargin=.5cm, roundcorner=2pt}

\definecolor{StringColor}{rgb}{0.06,0.10,0.98}
\definecolor{greencomments}{rgb}{0,0.5,0}
\definecolor{KWColor}{rgb}{0.37,0.08,0.25}
\definecolor{pantone288}{RGB}{1,31,91}
\definecolor{pantone201}{RGB}{153,0,0}
\definecolor{lightorange}{RGB}{255,247,230}

\lstdefinelanguage{assertLang}{
    basicstyle=\color{pantone288}\footnotesize\ttfamily,
    stepnumber=1,
    numbersep=8pt,
    string=[s]{"}{"},
    rulecolor=\color{black},
    keywordstyle=\color{KWColor}\bfseries,
    stringstyle=\color{StringColor},
    moredelim=[is][\textcolor{pantone201}]{\#}{\#},
    moredelim=[is][\textcolor{greencomments}]{\^}{\^},
    moredelim=*[is][\textcolor{scyan}]{\%}{\%},
    literate=
     *{(}{{\color{black}(}}{1}
      {)}{{\color{black})}}{1}
      {|}{{\color{black}|}}{1}
      {,}{{\color{black},}}{1}
      {@}{{\color{black}@}}{1}
      {!}{{\color{black}!}}{1}
      {[}{{\color{black}[}}{1}
      {]}{{\color{black}]}}{1}
      {\~}{{\color{KWColor}\textbf{$\sim$}}}{1},
}

\usepackage{enumitem}

\newenvironment{vinlist}
{\begin{itemize}[leftmargin=1.75em]
  \setlength{\itemsep}{0pt}
  \setlength{\parsep}{0pt}
  \setlength{\parskip}{2pt}
  \vspace{-4pt}
}
{\vspace{-4pt}\end{itemize}}

\newenvironment{vinenum}
{\begin{enumerate}[leftmargin=2em]
  \setlength{\itemsep}{0pt}
  \setlength{\parsep}{0pt}
  \setlength{\parskip}{2pt}
  \vspace{-4pt}
}
{\vspace{-4pt}\end{enumerate}}

\makeatletter
\renewcommand\subsubsection{\@startsection{subsubsection}{3}{\z@}%
 {-0.55\baselineskip \@plus -2\p@ \@minus -.2\p@}%
 {.4\baselineskip}%
 {\reset@font\bf}}
\makeatother

\begin{document}

\date{}

\title{\Large \bf FP4: Line-rate Greybox Fuzz Testing for P4 Switches \vspace{3mm}}
\author{
{\rm Nofel Yaseen}\\
University of Pennsylvania
\and
{\rm Liangcheng Yu}\\
University of Pennsylvania
\and
{\rm Caleb Stanford}\\
University of Pennsylvania
\and
{\rm Ryan Beckett}\\
Microsoft
\and
{\rm Vincent Liu}\\
University of Pennsylvania
} 

\maketitle

\begin{abstract}
Compared to fixed-function switches, the flexibility of programmable switches comes at a cost, as programmer mistakes frequently result in subtle bugs in the network data plane.

In this paper, we present the design and implementation of \name, a fuzz-testing framework for P4 switches that achieves high expressiveness, coverage, and scalability. \name directly tests running switches by generating semi-random input packets and observing their resulting execution in the data plane. 
%
To achieve high coverage and scalability, at runtime, \name leverages P4 itself with another ``tester'' switch that generates and mutates billions of test packets per second entirely in the dataplane. Because testing some program branches requires navigating complex semantic input requirements, \name additionally leverages the programmability of P4 by instrumenting the tested program to pass coverage information back to the tester through the packet header.

We present case studies showing that \name can validate both safety and stateful properties, improves efficiency over existing random packet generation baselines, and reaches 100\% coverage in under a minute on a wide range of examples.



\end{abstract}
\section{Introduction}
Computer networks have evolved to include more flexible platforms in which data plane functionality  can be defined by programmers using domain-specific languages like P4 that describe devices' data plane processing.
These devices are opening doors to better support applications \cite{sapio2020scaling, 10.1145/3132747.3132764} and to improve network management \cite{hira_wobker_2015}.
Although the increased programmability offers great benefits, it also brings the increased risk of introducing new bugs that are difficult to catch, either in the data plane, control plane, language compiler, ASIC implementation, or any combination of the above.

Bugs may arise as a result of anything from simple programmer typos to divergent interpretations of the P4 specification~\cite{10.1145/3434322}.
They can even include behavior that spans multiple packets because switches can store and recall state in registers, trigger control plane transitions, and reconfigure match-action rules as a result of incoming packets.
Combined, all of these factors increase the complexity of bugs in switches~\cite{10.1145/3230543.3230582,10.1145/3387514.3405888}.

To reduce bugs, recent work has suggested static verification to prove the correctness of P4 programs~\cite{10.1145/3387514.3405888, 10.1145/3230543.3230582, 10.1145/3185467.3185499}.
For pure, stateless data plane programs, static verification is often effective since there are no complex pointer-based data structures or loops, making analysis both more accurate and tractable.
However, real forwarding behavior depends on many other components: the control plane; the exact past, present and future match rules; the compiler translation; the switch state including registers; and specifics of the hardware implementation.
For example, in the process of developing \name, we discovered a subtle compiler/runtime bug in our installed SDE in which a multicast primitive in the default action of a table with no entries does not properly multicast the packet.
All other commands in the default action execute correctly as does the same setup in the provided simulator.
Static verification tools that only consider the P4 program itself cannot catch this class of bug.

We note that, in traditional programs, developers often rely on fuzz testing to catch this wider class of bugs.

A fuzzer generates semi-random inputs to discover assertion failures, memory leaks, and crashes.
Fuzz testing is able to evaluate applications in their natural environment (ignoring issues that are impossible to reach and catching issues that only arise in the presence of the application's surrounding components).
Fuzzing, and particularly blackbox and greybox fuzzing, also tends to scale well with the complexity of control flow and state.
As others have noted~\cite{aflfast}, these approaches often find more bugs than whitebox approaches like static verification and symbolic-execution-based test case generation as the latter either (a) spend significant time doing program analysis and constraint solving or (b) further sacrifice precision, e.g., by approximating functions that are hard to reason about analytically, such as hash functions.

In this work, we observe that, when applied to programmable switches, not only does fuzzing allow a developer to check the entire device \textit{in vivo}---incorporating effects of the data plane, control plane, ASIC implementation, and compiler---it also allows the fuzzer to leverage the intrinsic hardware parallelization, pipelining, and acceleration of packet processing in today's network devices.
Explicitly optimized for fast packet processing, switch-based fuzzing potentially enables input testing that is orders of magnitude faster than is possible in a CPU.

\begin{figure}[t]
    \centering
    \vspace{-3.8cm}
    \includegraphics[width=\columnwidth]{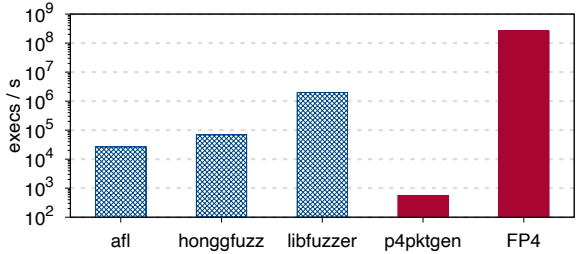}
    \vspace{-3.8cm}
    \caption{Maximum possible throughput of a single instance of modern fuzzing frameworks, both those for traditional programs (AFL, honggfuzz, and libfuzzer) and for P4 programs (p4pktgen and FP4).
    In each case, the fuzz target is an empty function or data plane program.}
    \label{fig:fuzz-perf}
\end{figure}

\definecolor{OliveGreen}{cmyk}{0.64,0,0.95,0.40}
\renewcommand{\checkmark}[0]{\textcolor{OliveGreen}{\ding{51}}}
\newcommand{\xmark}[0]{\textcolor{sred}{\ding{55}}}
\begin{table}[t]
    \small
    \setlength{\tabcolsep}{3pt}
    \begin{center}
        \begin{tabular}{ l c c c c c c }
            \toprule
             & \textbf{p4v} & \textbf{vera} & \textbf{p4wn} & \textbf{p4pktgen} & \textbf{P6} & \textbf{\name} \\ 
            \midrule
            \textit{type} & static & static & static & runtime & runtime & runtime \\
            \textit{line rate} & N/A & N/A & N/A & \xmark & \xmark & \checkmark \\
            \cmidrule(lr){1-7}
            \textit{DP logic} & \checkmark & \checkmark & \checkmark & \checkmark & \checkmark & \checkmark \\
            \textit{DP state} & \checkmark & \xmark & \checkmark & \xmark & \checkmark & \checkmark \\
            \textit{compiler} & \xmark & \xmark & \xmark & \checkmark & \checkmark & \checkmark \\
            \textit{control plane} & \xmark & \xmark & \xmark & \xmark & \xmark & \checkmark \\
            \textit{hardware} & \xmark & \xmark & \xmark & \xmark & \checkmark & \checkmark \\
            \bottomrule
        \end{tabular}
\caption{Comparison of the features of a selection of P4 verification and testing
    frameworks, including whether they can catch bugs in data-plane logic, stateful behavior, the compiler, the control-plane, and hardware.
    }
    \label{tab:comparison}
    \end{center}
\end{table}

To that end, we present \name, a greybox fuzz testing framework for P4-programmable network devices that is both (a) full-stack and (b) line-rate.
\name feeds semi-random packets to real programmable switches to attempt to trigger violations of programmer-specified assertions.
The switches are purposely kept as faithful as possible to their production deployments and run instrumented versions of their original P4 programs and control planes.

As a preview of the performance benefits of \name's approach, we measure executions per second for several traditional program fuzzers and p4pktgen, a software-emulated P4 fuzzer.
The results are shown in \cref{fig:fuzz-perf}.
%
%
All systems except \name were on an Intel Xeon E5-2660v3 2.60\,GHz CPU core with empty programs (empty parser and control block in the case of p4pktgen);
thus, these numbers represent an upper bound for prior work.
\name has two orders of magnitude higher throughput than the fastest traditional software fuzzer and almost 6 orders of magnitude faster than p4pktgen.

Unfortunately, implementing fuzz testing in P4 requires addressing numerous challenges.
First, to take advantage of the specialization of modern switch packet processing and achieve fast fuzzing speeds, we need methods to both generate and execute fuzzing at line rate.
Second, line rate packet input generation, on its own, is insufficient to find all bugs as the space of possible packets is large and often redundant.
In both traditional and programmable-switch fuzzing, careful choice of inputs is critical to good fuzzing performance.
Finally, typical methods to handle stateful behavior involve generating sequences of inputs and resetting the state between sequences (e.g.,~\cite{aflnet}).
Unfortunately, resetting switch state is a fundamentally expensive operation that would severely limit fuzzing performance. 

To address the first two challenges, \name takes inspiration from two different fuzz testing approaches: generator-based and coverage-guided fuzz testing. 
Generator-based fuzzers~\cite{10.1145/3293882.3330576} generate semi-random input such that inputs are passed through the input sanitation of the program.
\name knows the structure of the input from the headers and how they impact the processing of packet from the parser; and it uses it to generate valid packets. 
Coverage-guided fuzzing, on the other hand, leverages program instrumentation to trace the code coverage reached by each input and uses this information to make informed decisions about which inputs to mutate to maximize coverage.
\name tracks the actions visited by each packet in the dataplane by marking bits in the header.
Both are implemented, tracked, and learned quickly with the help of a second programmable switch.

To address the third challenge, \name splits switch state into a few categories.
For data-plane-only state, it leverages the fact that switches are meant to run continuously and, thus, most network tasks allow for intrinsic state resets (e.g., when a counter rolls over or a flow entry times out)~\cite{yaseen20aragog}.
Thus, continued fuzzing will eventually allow the switch to re-explore previous states. 
For everything else (i.e., control plane, table entry, or configuration state), \name borrows another idea from traditional fuzz testing---context-sensitive branch coverage~\cite{chen2018angora}---that seamlessly integrates state changes into greybox fuzzing approaches.



We implement and deploy \name to a hardware testbed in order to instrument and debug real P4 programs. 
\name works by modifying the input P4 program in a way such that it has no impact on normal packet processing. 
It also adds an extra header that stores information to track the actions visited and assertions failed by the packet. \name uses this tracking information to generate new seed packets.
Our work makes the following contributions:
\begin{vinlist}
    \item We leverage the observation that programmable switches can generate semi-random packets at line-rate to design, implement and evaluate fuzz testing framework for P4 programs that generates test packets 6 orders of magnitude faster than similar work for P4.
    \item We introduce a novel technique to instrument  P4 programs to track their coverage and check for assertion failures at line rate.
    \item We implement an \name prototype and evaluate it on a diverse set of P4 programs.
    Our results show \name achieves 100\% coverage quickly -- in <1\,min in most cases.
    \item To ensure reproducibility and facilitate future work, we will release \name as an open source tool on publication.
\end{vinlist}
\vspace{-1em}
\section{Background and Motivation}

In this section, we cover the challenges of discovering bugs in P4 switches and the possible role of fuzz testing.

\subsection{Potential Bugs in P4 Programs}


Bugs can occur in any point of the deployment and execution of a P4 program.
They can include but are not limited to:

\heading{Bugs in the application logic}
The most straightforward class of bugs exists in the P4 code itself.
In some cases, these issues are a result of ambiguities or subtleties in the language specification~\cite{10.1145/3434322}.
More generally, however, programmers are fundamentally fallible and just as capable of introducing bugs to P4 programs as they are to traditional code, especially when trying to reason about edge cases or complex interactions between features.
For example, a P4 reference program previously contained a bug where ACL rules were incorrectly applied to control-plane traffic~\cite{switch-p4-bug}.

\heading{Issues in the compiler or hardware implementation}
The P4 program must be compiled to run on the hardware and optimized to adhere to resource limitations on pipeline stages, SRAM, TCAM, etc.
As above, the programmers of these components are also fallible, creating instances where an otherwise correct P4 program produces unexpected behavior.
While this class of bugs is typically rarer due to longer development cycles, lower-level specifications, and heavier testing, the above multicast issue and a glance at the errata of any processor or compiler documentation validates their presence.
These are among the most difficult type of error to diagnose.

\heading{Bugs in the control plane}
Switch operation depends on the combination of the data and control planes.
While the data plane is responsible for handling per-packet processing, the control plane---operating in parallel on a general-purpose CPU---is responsible for managing the data plane and handling all of the tasks that are too complex for line-rate processing.
These include installing, updating, and removing data-plane table rules as well as executing routing protocols.
All of these can evolve based on the sequence of incoming packets---real control planes are both dynamic and stateful.

\heading{Switch misconfigurations}
Finally, network switch behavior is also affected by switch configuration options like knobs in the traffic manager, buffer slicing, and port speeds.
These configuration options can be independent and set separate from either the traditional data plane or control plane programs.
Errors can arise either from operator misconfiguration or through interactions with other issues, e.g., in the hardware implementation.

\noheading
Bugs can occur within any of the above components, and some only manifest when issues in multiple layers combine.
\subsection{Fuzz Testing}

For traditional applications, programmers often augment their software engineering workflows with fuzz testing~\cite{oss-fuzz}.
Fuzz testing feeds the program a set of random inputs and observes whether the program behaves correctly on each such input.
This process is able to automatically discover bugs, even when those bugs result from complex runtime behavior and interactions between heterogenous systems.


As prior work has noted, however, the \naive approach of random inputs (i.e., pure blackbox fuzzing) can often lead to poor coverage as many inputs are simply invalid or fail to explore program paths with complex or hard-to-hit branch conditions~\cite{dart-testing}.
On the other hand, approaches that try to reason precisely about the program's structure (i.e., pure whitebox fuzzers) come with their own set of issues ranging from being unable to model complex functions (e.g., hash functions) to exhibiting poor scaling that makes them not worth the extra overhead~\cite{aflfast}.
In the end, many of the most prolific fuzzers take a greybox approach that attempts to strike a balance.
\name's input generation takes inspiration from three methods from the literature on greybox fuzzing of traditional programs:

\headingraw{(1) Coverage-guided fuzzing} is exemplified by the widely used AFL fuzzer~\cite{afl-fuzz}.
AFL begins with a set of seed inputs that it subsequently mutates to create random inputs that are then fed to the program. 
Based on feedback from the program (detailing paths and branches covered), AFL learns the quality of the inputs and selectively updates the set of seed inputs to explore new execution paths.
This leads to significant improvement in the rate of coverage compared to completely random inputs.

\headingraw{(2) Generator-based fuzz testing} allows users to write generator programs for producing inputs (see~\cite{10.1145/3293882.3330576}).
As an example, consider the structure of an Ethernet/IP protocol stack, which might accept IP-related EtherTypes and discard all other packets as corrupted or otherwise invalid.
A generator-based fuzzer will generate only valid IP packet inputs.
This ensures that the fuzzer does not waste time on inputs that are immediately discarded by input sanitation.

\headingraw{(3) Context-sensitive branch coverage} is introduced in the Angora fuzzer (see~\cite{chen2018angora}, Section 3.2). Angora observes that not every execution of the same code block
(containing a conditional branch)
is equal.
Instead, the current state of the program and its call stack
can make an execution of the same code block materially different from prior executions.
Including this context in coverage tracking therefore improves feedback and responsiveness.

\begin{figure*}
    \centering
    \includegraphics[width=.7\linewidth]{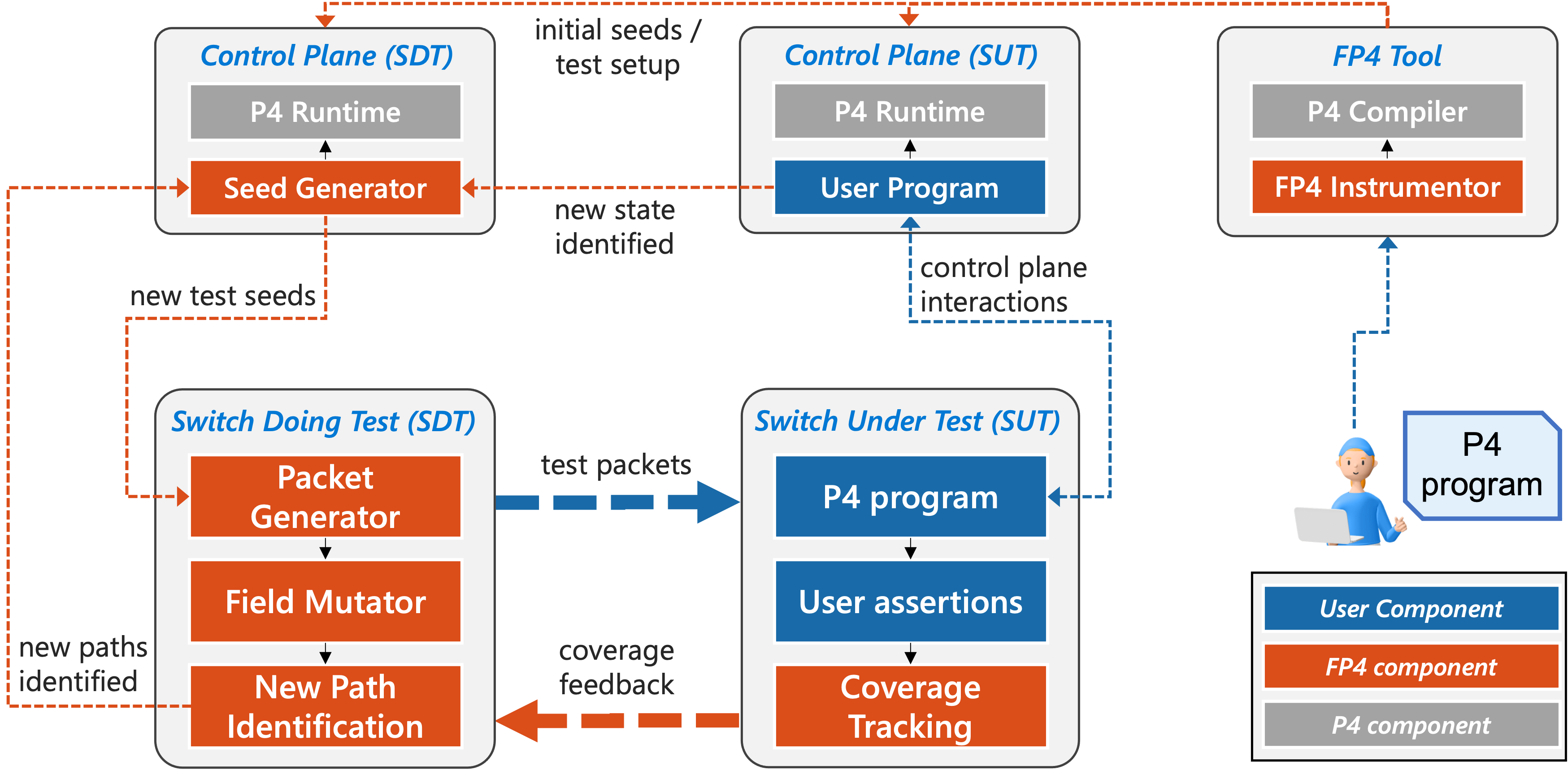}
    \vspace{.6em}
    \caption{System design of \name. An operator writes assertions in the P4 program.
    The program is an input to (1) instrumentation (Section~\ref{sec:sut}) that adds statements to track packets and (2) program synthesizer (Section~\ref{sec:sdt}) that generates the P4 program and control plane to conduct the test. After installing respective programs on both switches, \name runs fuzz testing. \name generates valid packets (generator based fuzzing) and adds new seed packets based on coverage information (coverage guided fuzzing).
    }
    \label{fig:system}
\end{figure*}

\section{Overview}


Like most other coverage-guided greybox fuzzers,
\name is built around a single loop in which \name generates a packet from a selected set of seeds, mutates the packet semi-randomly, passes it through the target (programmable switch), and computes the state and path coverage to determine whether the packet is a good candidate for a new seed.
However, unlike other fuzzers, \name is extremely fast.
It achieves this speed with two domain-specific insights:

\begin{vinenum}
\item Modern switches can execute packet processing orders of magnitude faster than commodity CPUs, and that speed is independent of the complexity of the program as long as it fits within a single pass through the switch.
As mentioned, running the target system in vivo provides benefits to the speed, completeness, and accuracy of fuzz testing. 
A key contribution of \name is to demonstrate that the fuzzer is \textit{also} deployable to programmable switches, generating, modifying, and checking at line rate.

\item Switch programs are intended to be long-lived.
This means that, in steady state, switch programs typically have intrinsic mechanisms that reset persistent state, e.g., when a counter overflows, when a ring buffer wraps, or when routes are torn down.
This allows \name
to test most state transitions without needing an explicit reset of the switch.
An exception are bugs that occur during initialization, but those are typically straightforward for operators to catch during development and canarying.
\end{vinenum}



An \name testbed consists of two switches: (1) the switch under test, which runs the target data plane, control plane, and switch configuration; and (2) the switch doing the test, which generates inputs, checks program coverage, and manages seed packet selection.
See \cref{fig:system} for a visual depiction.


\heading{Switch Under Test (SUT)}
The SUT executes the target switch system, including both its P4 data plane and control plane program.
The system should operate identically to a real deployment with the exception of some additional annotations and instrumentation.

The annotations come in the form of a simple \textit{operator-specified error conditions} on a given packet's contents or the state in the switch (i.e., registers, counters, and meters).

In the automatic \textit{instrumentation step}, \name then inserts code into both the data plane and (optionally) control plane programs to aid in checking for path coverage to trigger the above errors.
This instrumentation takes the form of an additional packet header, an operation in every data-plane action, emulated output ports, and a couple of additional tables for bookkeeping and assertion checking.
All of the above changes incur minimal overhead and, crucially, leave the original metadata, headers, and control path intact.



\heading{Switch Doing Test (SDT)}
Alongside the SUT, we run a second switch, the SDT.
The SDT is responsible for all of the traditional tasks of a fuzzer: generating test packets, mutating them, sending them to the SUT, and checking for violations and coverage after packets return from the SUT.

It consists of a \textit{data-plane test generator}, which is a synthesized companion P4 program that generates the test packets, mutates them, tracks coverage, and checks for assertion failures.
The generator leverages a set of dynamically updated seed packets to generate billions of semi-random test packets per second.
\name mutates the packets at line rate using the programmability of the P4 switch.
The mutations are such that they retain the validity of the packet but attempt to steadily increase the coverage of the fuzz testing.

Supporting the data-plane test generator is a \textit{control-plane fuzzing manager} that is used to consider seed packets candidates, modify the data-plane generator accordingly, notify users of assertion failures, and perform other tasks that are beyond the capabilities of today's programmable data planes.
Like in traditional networks, executing these tasks asynchronously in the control-plane CPU (while carefully maintaining correctness) allows the data plane to continue operating at line rate.

\noheading
The remainder of this paper describes the design of \name's SUT and SDT in more detail.

\color{black}
\section{Switch Under Test (SUT) Instrumentation} \label{sec:sut}

We begin by outlining \name's modifications to the SUT.

\begin{figure}[!t]
\begin{mdframed}[style=background]
\begin{lstlisting}[language=P4,xleftmargin=1.5ex,xrightmargin=1.5ex]
header_type #fp4_header_t# {
    fields {
        visited_action1 : %1%;
        visited_action2 : %1%;
        ...
        assertion1 : %1%;
        ...
        // 0 -> freshly generated packet
        // 1 -> additional mutation needed
        // 2 -> completed packet
        // 3 -> state change from SUT control plane
        pkt_typ : %2%;
}}
\end{lstlisting}
\end{mdframed}
\caption{Structure of the \name header. 
    }
\label{lst:fp4_header}
\end{figure}

\subsection{Programmer Assertions}

One type of instrumentation in \name involves programmers adding assertions in their P4 programs that define error conditions in the processing of a packet.
One example of violation might be where the time to live field of a packet is zero, but the program fails to actually drop the packet:
\begin{alltt}\small
   assert(ipv4.ttl != 0 || std_metadata.drop == 1)
\end{alltt}
More generally, operators specify fields and their range of invalid values using basic comparison operators and boolean logic. 

These assertions serve as syntactic sugar that \name uses to automatically generate a set of tables, actions, and table rules that will catch the assertion at runtime.
Violations are marked in an \name packet header that is appended to the packet (see \cref{lst:fp4_header} for the header's format).
Note that operators can use this syntax to detect issues that span multiple packets by manually tracking relevant information in stateful elements.

\subsection{Coverage Instrumentation}
\label{sec:instrumentation}

The other type of instrumentation in \name enables its greybox, coverage-guided fuzzing within the data plane.
Traditional fuzzers typically track coverage at the granularity of basic blocks, adding instrumentation to each branch to record `seed-worthy' inputs that trigger additional program coverage (i.e., that are not redundant with existing seeds).
Programmable switches, with their concomitant control planes and frequent rearrangements of control flow (via control plane intervention), impose additional restrictions on what it means for an input to be worthy of use as a seed.
\name considers:

\begin{vinlist}
    \item \textit{Actions:} In most P4 implementations, the most convenient single-entry, single-exit, straight-line (with the exception of ALU operations) block of code are the actions of the match-action pipeline. 
    The goal of \name is to fuzz test all possible actions, so coverage of novel actions is cause for addition to the input corpus.

    \item \textit{Table entries:} The actions that are triggered and the conditions under which they are triggered are determined by the table entries of the match-action pipeline.
    The overall path is defined by a sequence of table entry hits.
    Thus, table entries---and in particular, their union---have a massive influence on the reachability of bugs.
    
    \item \textit{Control plane state:}
    Finally, while \name, its design, and its assertions are primarily focused on bugs in the data plane, we note that packets sometimes pass through the control plane as part of their processing (e.g., routing updates that eventually add/remove table entries).
    \name is not concerned with the code coverage of the control-plane program but does care about how it might eventually affect the data plane, e.g., through table entry updates.
    
\end{vinlist}

Purposefully missing from the above set is data-plane state.
While data-plane state (like table entries and control plane state) may also impact control flow and lead to additional program coverage, we found that properly tracking the uniqueness of data-plane state in the presence of per-packet register access limitations and packet reordering imposed too much overhead and too many limitations on the scope of P4 programs that \name can test.
We leave an exploration of more efficient methods of state tracking to future work.

In the remainder of this section, we describe the SUT instrumentation required to track changes to the above entities.

\subsubsection{Actions Visited}

To track the coverage of every action, \name assigns a bit in the \code{fp4\_header} for each action, and marks the respective bit in each packet as it passes through the switch pipeline.
For example, consider the target program of \cref{lst:running_example}, there are four total actions and, thus, four reserved bits in the header.
Laying it out in this way ensures that every unique path through the pipeline results corresponds with a unique `visited' bitstring.

\begin{figure}[!t]
\centering
\begin{mdframed}[style=background]
\begin{lstlisting}[language=P4,xleftmargin=2ex,xrightmargin=2ex,basicstyle=\scriptsize\ttfamily\color{sbase00}]
parser #parse_ethernet# {
    extract(ethernet);
    return select(latest.etherType) {
        %ETHERTYPE_IPV4% : #parse_ipv4#;
        default: #ingress#;
}}
parser #parse_ipv4# {
    extract(ipv4);
    return #ingress#;
}
action #on_l2_hit#(vrf){
    #modify_field#(l3_metadata.vrf, vrf);
}
table #ethernet_forward# {
    reads { ethernet.dstAddr : exact; }
    actions { #on_l2_hit#; #on_l2_miss#; }
}
table #ipv4_forward# {
    reads { l3_metadata.vrf : exact;
             ipv4.dstAddr : lpm; }
    actions { #on_l3_hit#; #on_l3_miss#; }
}
control #ingress# {
    apply(#ethernet_forward#);
    if #valid#(ipv4) { apply(#ipv4_forward#); }
}
\end{lstlisting}
\end{mdframed}
\caption{A simple example target P4 program.}
\label{lst:running_example}
\end{figure}

Note that if the same action is used in multiple tables, a \naive application of the above may leave ambiguity in the packet's path through the processing pipeline.
\name addresses this by duplicating the action and renaming it, so each action is unique to a table. 
The renaming has no impact on the switch hardware resources.

\subsubsection{Control-plane State Changes}
\label{sec:sut_control_instrumentation}

To account for the impact of control-plane changes, \name augments the control plane to track its internal state.
Note that this can include everything relevant to the processing of future packets, from object attributes that persist across packet events to the current state of the stack (which reflects function calls, parameter changes, and returns). This additional data can only improve coverage, but is not necessary for functionality.

\Naive{}ly, one could consider every packet that causes a state change as a candidate for inclusion in the seed corpus.
Unfortunately,
this fails to distinguish new states from previously seen ones.
Instead, \name leverages the CRC-32 algorithm to compute a 32-bit hash of the control plane state (all global variables followed by the sequence of function calls on the stack)
that is both efficient to update and can distinguish between unique states.
CRC-32 values can be updated bi-directionally (i.e., they support both pushing and popping bytes), so the hash is maintained on both function calls and returns in constant time.
Our \name prototype implements this approach with a semi-automatic annotation process.
It currently assumes that the control plane is written in Python and all functionality is contained within a single class.
Programmers annotate the class with a superclass 
and add a decorator to each of its methods and local variables, which wraps them to update the CRC value on each call, return, and modification (in principle, these can be automated).
\name also automatically stores the last-seen packet/digest from the data plane and wraps all table entry modifications.


Whenever a table entry is added or the control-plane state changes, \name checks if the CRC value is novel and there is an active `input packet/digest.'
If both are true, the SUT control plane forwards the new state hash and the original headers of the input packet to the SDT for inclusion in the seed corpus.

\subsubsection{Table Entry and Configuration Changes}

Runtime updates to table entries and switch configurations can also affect the data-plane behavior, and \name tracks them using a similar technique as above.
Specifically, \name automatically computes the CRC-32 of the string representation of all the configuration changes (the value is kept separate from the hash of internal control-plane state).
For example, it interposes on the table write/update library calls to automatically track the content of every table.
It represents each entry in its runtime-CLI-command format (which provides a simple, unique representation of the entry), and computes the hash of a sorted list of such commands.
As above, \name automatically checks the uniqueness of the state and, if unique, forwards the input packet to the SDT.



\section{Switch Doing Test (SDT) Design}
\label{sec:sdt}

\begin{figure}
    \centering
    \includegraphics[width=\linewidth]{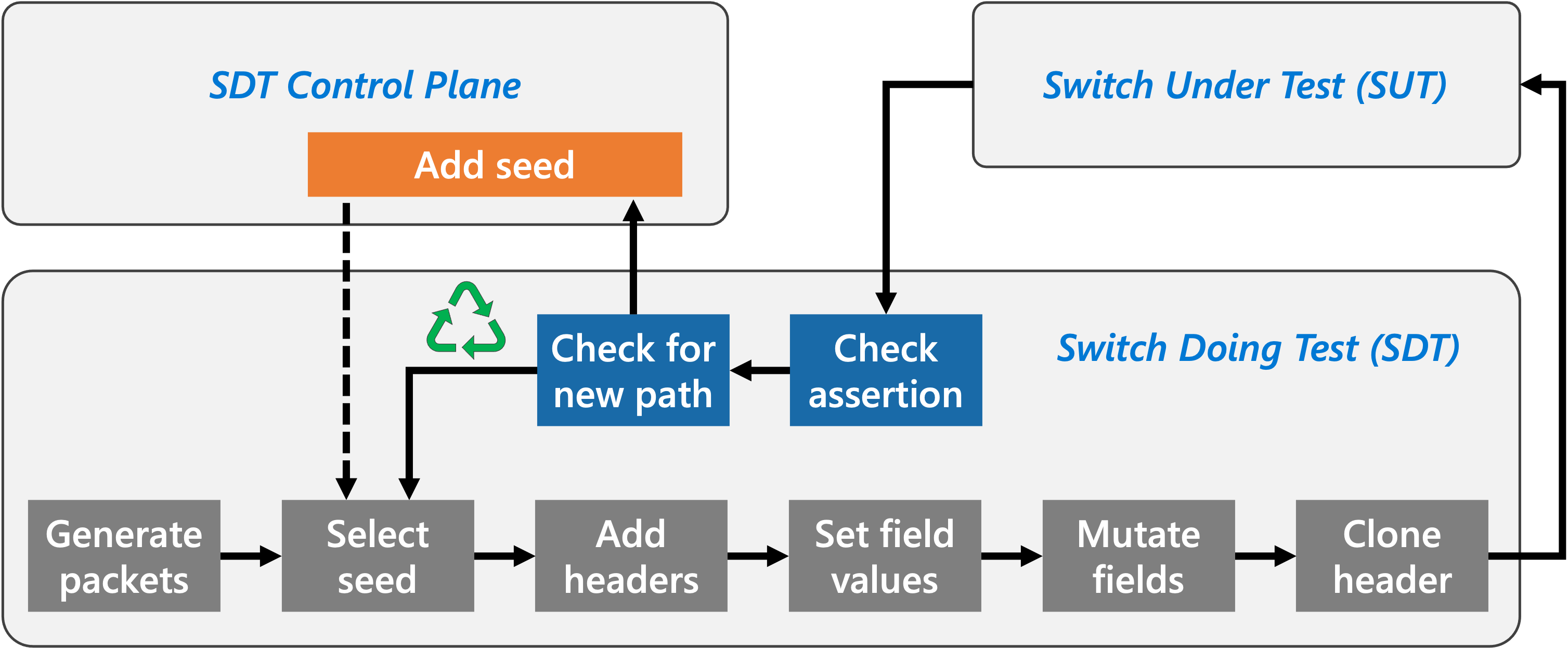}
    \caption{Lifecycle of a packet in \name.
    }
    \label{fig:packet_lifecycle}
\end{figure}

To handle the fuzzer tasks, \name leverages a second programmable switch: the SDT.
For the tasks that must execute for \textit{every} fuzzing input, \name leverages the line-rate processing capabilities of the SDT's data plane.
For other tasks, \name leverages the SDT's control-plane CPU to implement more complex behaviors that improves its choice of inputs.
In total, the SDT generates test packets, mutates them, sends them to the SUT, and checks for violations/coverage after packets return from the SUT. \cref{fig:packet_lifecycle} illustrates this lifecycle.

\subsection{Packet Generation} \label{sec:seed_packet}

To generate the input packets for the target at line rate, \name's SDT uses the built-in hardware packet generation capabilities of the Tofino and similar switches.
Generated packets contain an \code{fp4\_visited} header (depicted in \cref{lst:fp4_header}) with all fields initialized to 0s.
The generated packets are then passed through automatically synthesized tables that transform the zeroed packet into one of a limited set of seed packets.
The tables first select a random seed number.
Based on that seed, the SDT will add a set of headers to the packet and fill them in with an initial value corresponding to one of the configured seed packets.
\cref{lst:ti_create_packet} provides a snippet of the synthesized actions that \name uses to transform the generated packets.

When \name is first executed, it only contains few seed packets based on the program in SUT; more are added during runtime (see \cref{sec:sdt_control}).

\heading{Deriving expected packet formats}
The first step in any P4 program is the parser, which takes a sequence of bits from the MAC layer and parses it into its constituent headers.
Packets at this stage must adhere to strict formats---all others are dropped before reaching the ingress pipeline of the switch.
When generating packets, \name ensures that all seeds (whether from the initial set or added later) pass this stage using a generator-based fuzzing approach.

More specifically, we note that P4 parsers are structured as state machines.
The parser transitions between different states depending on the contents of the packet;
a packet is only fed into the packet processing pipeline if it reaches an accept state in the parser.
\name analyzes the state machine to find all paths from the start to any terminal state; it records the headers extracted in each path along with any field and header contents that triggered the path.

\name synthesizes implementations of the above seed-packet generation tables so that it is able to configure seed packet contents from the control plane.
Adding a seed is as simple as inserting a table rule to the above tables.

\heading{Computing an initial set of seeds}
\name also uses information from the parser to compute the initial set of seeds. 
Specifically, it pre-computes one seed packet for each unique parser path,
setting the content of the seed packet header to specific values that are part of transitions in the state machine%
\footnote{Note that, like other implementations of the P4 compiler, \name limits parser recursion to a specified depth, ensuring finite seed packets.}%
.
All other fields in the seed header that are \textit{not} constrained by the parser state machine transitions are randomly populated.




As an example, consider \cref{lst:running_example} and its synthesized actions in \cref{lst:ti_create_packet}.
The parser for this program always extracts an Ethernet header, and then only extracts an IPv4 header if the contents of the EtherType are ``0x0800.'' 
Its state machine, therefore, consists of three states: the start state (not shown), a state to parse the Ethernet header, and a state to parse the IPv4 header.
Further, there are only two unique paths through this state machine that lead to accepting states: (1) an L2 frame with only an Ethernet header and (2) an L3 packet with both Ethernet and IPv4 headers.
During initialization of the SDT, \name discovers these two paths and randomly generates two initial seed packets that will trigger these parser paths.

\begin{figure}[!t]

\begin{mdframed}[style=background]
\begin{lstlisting}[language=P4,xleftmargin=1.5ex,xrightmargin=1.5ex]
action #add_ethernet_ipv4_header#() {
    #add_header#(ethernet);
    #add_header#(ipv4);
    #add_header#(ethernet_original);
    #add_header#(ipv4_original);
    #modify_field#(ethernet.etherType, %0x0800%);
}

action #add_ethernet_ipv4_content#(ethdstAddr, ...) {
    #modify_field#(ethernet.dstAddr, ethdstAddr);
    #modify_field#(ethernet.srcAddr, ethsrcAddr);
    #modify_field#(ipv4.version, ipv4version);
    ...
}
\end{lstlisting}
\end{mdframed}
\caption{Actions that are used to create an Ethernet+IPv4 packet from an existing seed.
These correspond to the `Add headers' and `Set field values' steps of \cref{fig:packet_lifecycle}.
For the example target of \cref{lst:running_example}, a separate set of actions would be synthesized for creating Ethernet-only packets.}
\label{lst:ti_create_packet}
\end{figure}




\subsection{Mutating the Generated Packets}
\label{sec:sdt_mutations}


With the seed packet in hand, \name's goal is then to use the packet to expose new switch behavior.
A simple straw man approach to mutating packets would be to randomly select a subset of the header fields and set them to random values.
While such an approach will \textit{eventually} catch any bug, blindly mutating packets may only rarely result in inputs that traverse new control flow or trigger new table actions.
Instead, \name takes a more targeted approach using one of three techniques per packet (with configurable probability of each decision).

\heading{(1) Targeting a specific \textit{table entry} or \textit{conditional statement}}
Fundamentally, code coverage is determined by the actions triggered in the SUT.
For a packet to trigger a given action, its headers and metadata must match the set of `keys' in a table entry that is mapped to the target action.
\name takes advantage of the fact that the current set of table entries are known precisely at runtime to implement a `magic value' approach to fuzzing~\cite{rawat2017vuzzer,li2017steelix}.

More specifically, the SDT contains a `mutation' table with an action corresponding to each match-action table and conditional statement in the SUT.
In the case of a SUT table, the action takes its match fields as parameters;
thus, whenever an entry is added to a SUT table, \name can attempt to add a corresponding entry to the mutation table with the match-key constants passed as parameters to the table-specific action.
LPM keys are converted to their base value; `do not care' bits of ternary matches are converted to zeros.
In the case of a conditional, the action takes any referenced fields as parameters, and \name tries to add a corresponding entry based on static analysis of the program.


When target values are sparse and directly dependent on the input packet header, this technique can greatly speed coverage.
For example, consider an ingress MAC filter that only matches the interface and broadcast MAC addresses, two values out of $2^{48}$.
Even at $\sim$2 billion packets per second, triggering the action would take an average of $\sim$20 hours.


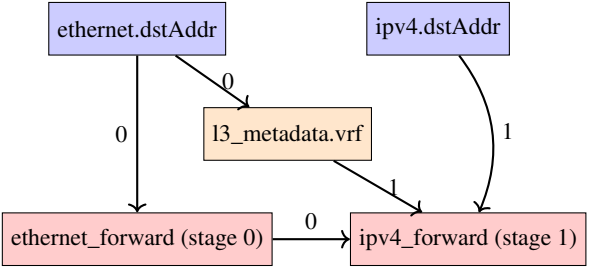
\begin{figure}
\centering
\begin{tikzpicture}
    \node[rectangle,draw,fill=blue!20,minimum height=.7cm,minimum width=1.8cm] (f1) at (0,0) {\small ethernet.dstAddr};
    \node[rectangle,draw,fill=blue!20,minimum height=.7cm,minimum width=1.8cm] (f3) at (4,0) {\small ipv4.dstAddr};
    \node[rectangle,draw,fill=orange!20,minimum height=.7cm,minimum width=1.8cm] (m) at (2,-1.4) {\small l3\_metadata.vrf};
    \node[rectangle,draw,fill=red!20,minimum height=.7cm,minimum width=1.8cm] (t1) at (0,-2.8) {\small ethernet\_forward (stage 0)};
    \node[rectangle,draw,fill=red!20,minimum height=.7cm,minimum width=1.8cm] (t2) at (4.4,-2.8) {\small ipv4\_forward (stage 1)};
    \draw [draw,->,thick,right] (f1) -- node {\small 0} (m);
    \draw [draw,->,thick,left] (f1) -- node {\small 0} (t1);
    \draw [draw,->,thick,right] (m) -- node {\small 1} (t2);
    \draw [draw,->,thick,right] (f3) to[bend left] node {\small 1} (t2);
    \draw [draw,->,thick,above] (t1) -- node {\small 0} (t2);
\end{tikzpicture}
\caption{Example dependency graph for the example of \cref{lst:running_example}.
Arrows indicate a ``depends on'' relation where the source node affects the computation of the target node.
Tables are marked with their stage, and arrows are labelled with the earliest stage with the dependency.
A table depends on a field if there is a path from the field to the table with only edges that have labels less than or equal to the table's stage.
}
\label{fig:mutate_example}
\end{figure}

\heading{(2) Targeting a \textit{table} or \textit{if statement}}
Note that not all matched fields are directly configurable.
Examples include tables that match on metadata or on fields that are changed in previous stages of the pipeline.
For these cases, \name takes advantage of the P4 program's structure to preferentially mutate a set of fields together if it knows these fields are more likely to result in ``hitting'' a new table entry or conditional branch, thus dramatically shrinking the search space of mutations.

But how does \name decide what groups of fields should be mutated together? To determine this, \name makes use of a lightweight, stage-sensitive static analysis over the program control flow graph (CFG).
The analysis extends traditional flow-insensitive static analysis techniques~\cite{muchnick1997advanced}
to be sensitive to packet modifications occurring at different stages.
In particular, \name statically analyzes the input program to create a dependency graph that captures whether each packet field ``could be'' relevant for a given table lookup.
The graph for \cref{lst:running_example} is shown in \cref{fig:mutate_example}.
It contains nodes for each of the packet's header and metadata fields as well as for each table appearing in the program. 

In P4, each table is associated with a stage number.
\name labels the table nodes with that stage number and adds an arrow between two nodes when there is a dependency between these components.
For example, if \code{on\_l3\_hit()} modifies the \code{egress\_port} of the packet, the \code{egress\_port} field would depend on the content of both the \code{ipv4.dstAddr} and a \code{vrf} metadata field.
The \code{vrf} field may, in turn, depend on \code{ethernet.dstAddr} if it is modified in an action of the \code{ethernet\_forward} table.
\name adds arrows for fields that are used as keys in tables as well as between tables whose actions may influence the future lookup of other tables.
It also adds dependency edges when fields are used in conditionals or updates of stateful ALUs.
In most cases, there is only one edge between tables;
however, when an \code{if} statement immediately follows a table lookup, there can be more than one dependency edge as there are multiple possible next tables.

Each dependency edge is labelled with the earliest stage in which the dependency occurs.
The mutation procedure is, thus, as follows:
For each table in the graph, \name precomputes the set of all packet fields that can impact that table. These fields are all those that can reach the table node in the dependency graph using only edges with stage labels less than or equal to the table stage.
At runtime, for each seed packet, \name stores the list of tables that this seed might be able to mutate given the fields present in the seed (which might be a subset of all possible fields).
During mutation, after selecting seed packet, it picks one table from the list and preferentially mutate together fields that are keys of that table.
%
%


\heading{(3) Targeting \textit{stateful counters}}
Finally, while \name does not explicitly track data-plane state (for the reasons in \cref{sec:instrumentation}), \name can still discover bugs that depend on stateful behavior.
The table-targeted mutation technique, for instance, will properly track the dependencies of stateful registers and modify them if the register outputs are useful for increasing coverage or testing assertions (it will not attempt to mutate state that is write-only).
There are, however, edge cases where \name's lack of visibility into data-plane state changes may impede its efficiency.
For example, consider a target program that counts the number of packets for each 5-tuple and triggers an assertion violation if any counter exceeds a threshold.
\name will quickly cover the action with the stateful counter, and it will eventually trigger the violation (after sufficient random collisions), but may do so slowly.
Noticing this tendency toward counters in network programs, \name will  occasionally repeat packets to ensure that the most common classes of stateful behaviors are captured quickly.


\heading{Chaining mutations}
Note that, particularly for (1) and (2), it may be advantageous to chain mutations to trigger matches that are impossible with only one mutation of existing seeds.
\name can pack a few such mutations within a single pipeline.
It can also optionally recirculate the packet to apply even more rounds of mutations, albeit at the cost of throughput.

\subsection{Evaluating Assertions and Coverage}
\label{sec:sdt_control}

At the end of pipeline, \name makes a copy of all headers to reserved \code{*\_original} headers.
If the packet is later determined to be seed-worthy, the cloned headers serve as a record of the original input packet, prior to any SUT modifications.

Two types of packets will return from the SUT: test packets that have traversed the SUT data plane and state-change notifications from the SUT control plane.
It may also receive packets generated at the SUT and destined for remote devices (e.g., a periodic control plane routing keepalive message), but these never result in an addition of seed packet (as they are not the result of an SDT input).

For packets from the SUT, the header will contain the visited action bitstring and assertion failure flags along with the original header.
The packet has attained additional coverage iff its visited bitstring is novel, i.e., it visited a unique sequence of actions or path.
Because the string can potentially be large, \name tracks uniqueness with the help of a bloom filter.
On a filter miss or a set assertion flag, \name sends the packet and its original header to the SDT control plane for further processing.
Otherwise, \name recycles the packet by removing all the headers;
the recycled packets are treated as a freshly generated packet.
Packets from the SUT control plane are always sent to the SDT control plane for addition to the seed packet set.
The SUT control plane should have already verified its uniqueness.

\section{Implementation}
\label{sec:implementation}

We implemented a prototype of \name, including the SUT instrumentation and SDT data plane and control plane agent.
Our hardware testbed consists of two Barefoot Wedge100BF-32X programmable switches with all ports on both switches connected by an array of 100\,GbE DAC cables.
Our implementation currently uses a single line card on each switch.

\heading{SUT Instrumentation} 
The \name instrumentation adds the required changes to the input P4 program stated in \cref{sec:sut} to generate an instrumented P4 program. 
In total, instrumentation implementation comprises around 5500 lines of C++ code, with 4300 lines for a frontend to parse input P4 code using Flex/Bison and build an AST of the input program and 1200 lines to instrument the target SUT program.

The SUT control plane scaffolding currently requires that the programmer annotate their code as described in \cref{sec:sut_control_instrumentation} and adhere to a general entrypoint signature.

\heading{SDT implementation}
Our prototype SDT data plane implements the pipeline and functionality detailed in \cref{sec:sdt}.
The code to synthesize the program for SDT incorporates an additional 4200 lines of code.
The program synthesis code also outputs a json file to be used by the SDT control plane. 
This json file contains the structure of packets so the control-plane can parse the incoming packets.

Hardware limitations in the per-stage random number generator restrict the size of mutations to 32-bits, but we find that table-entry-targeted mutations are sufficient to fill this gap.
Our prototype SDT control plane takes in the json file generated during instrumentation, adds initial seed packets, parses packets coming from the data plane, track coverage and installs new seed packets in the SDT data plane. 
The Python control plane is more than 1200 lines of code.

\section{Evaluation}

\begin{table*}[!t]
\centering
\footnotesize
\setlength{\tabcolsep}{4pt}
\begin{tabularx}{\textwidth}{ X c c c c r r r r r}
\toprule
\textbf{Program}
& \multicolumn{4}{c}{\textbf{Features}}
& \multicolumn{4}{c}{\textbf{Resource Overhead}}
& \multicolumn{1}{c}{\textbf{Coverage}} \\
\cmidrule(lr){2-5}\cmidrule(lr){6-9}\cmidrule(lr){10-10}
& {LoC} & {Actions} & {Stateful ALUs} & \multicolumn{1}{c}{Control Plane}
& \multicolumn{1}{c}{Stages} & \multicolumn{1}{c}{Tables} & \multicolumn{1}{c}{SRAM (KB)} & \multicolumn{1}{c}{Metadata (b)}
& \multicolumn{1}{c}{(s)} \\
\midrule
Load Balancer
& 159 & 4 & 0 & Static
& 2$\rightarrow$3 (+1) & 4$\rightarrow$6 (+2) & 144$\rightarrow$160 (+16) & 948$\rightarrow$1056 (+108)
& 0.79 (100\%) \\
Basic Routing
& 165 & 6 & 0 & Static
& 6$\rightarrow$8 (+2) & 9$\rightarrow$11 (+2) & 2080$\rightarrow$2096 (+16) & 647 $\rightarrow$ 747 (+70)
& 54.33 (100\%) \\
Rate Limiter
& 197 & 7 & 3 & Static
& 5$\rightarrow$6 (+1) & 7$\rightarrow$11 (+4) & 128$\rightarrow$144 (+16) & 755$\rightarrow$879 (+124)
& 8.53 (100\%) \\
Firewall
& 313 & 12 & 4 & Static
& 7$\rightarrow$8 (+1) & 12$\rightarrow$18 (+6) & 160$\rightarrow$176 (+16) & 1043$\rightarrow$1178 (+135)
& 1.66 (100\%) \\
Netchain~\cite{jin2018netchain}
& 264 & 6 & 2 & Static
& 3$\rightarrow$6 (+3) & 6$\rightarrow$10 (+4) & 544$\rightarrow$560 (+16) & 1084$\rightarrow$1203 (+119)
& 1.11 (100\%) \\
Mirroring & 213 & 7 & 4 & Static
& 1$\rightarrow$4 (+3) & 7$\rightarrow$13 (+6) & 128$\rightarrow$128 \phantom{(+16)}\makebox[0pt][r]{(+0)} & 651$\rightarrow$781 (+130)
& 0.24 (100\%) \\
DV Router & 284 & 11 & 0 & Dynamic
& 3$\rightarrow$4 (+1) & 11$\rightarrow$15 (+4) & 352$\rightarrow$400 (+48) & 949$\rightarrow$1104 (+155)
& 39.14 \phantom{(100\%)}\makebox[0pt][r]{(93\%)}
\\

\bottomrule
\end{tabularx}
\caption{P4 programs on which we evaluate \name, resource overhead of the instrumentation, and time to achieve full coverage.
}
\label{tab:test_programs}
\label{tab:resources}
\end{table*}

To evaluate the performance of \name, we conducted experiments on a diverse collection of programs that vary in size and complexity.
Rather than merely reaching a particular behavior (such as an assertion violation or invalid header access on a particular line), we focus on the more holistic problem of achieving full \emph{coverage} of all actions and paths in P4 programs, which can be combined with assertions to catch specific bugs.
As such, our evaluation aims to address the following questions: (1) How quickly does \name{} achieve 100\% code coverage, compared to existing tools for software-based fuzz testing?
(2) Which factors of \name{}s design have the biggest impact on coverage, and how does its performance compare to more \naive{} baselines? 
(3) What is the performance overhead of the added \name{} instrumentation on the switch under test?  (4) Finally, can \name{} be used successfully to find bugs in existing P4 programs---with a particular eye to bugs that could not be caught with static verification techniques?

\heading{Programs tested}
\cref{tab:test_programs} lists tested programs.
Load Balancer, Basic Routing, and Rate Limiter represent programs designed primarily for packet forwarding.
The Load Balancer makes use of hashing, while Rate Limiter tests Stateful ALUs.
Firewall represents a more complex application using Bloom filtering. Netchain also uses concurrency control.

In order to avoid false positives due to invalid table rules inserted by the control plane~\cite{10.1145/3387514.3405888}, for these examples, we employ a static controller which inserts a fixed number of table rules.

We also evaluate an additional two programs, Mirroring and DV Router.
The first program contains the bug mentioned in the introduction: our goal is to determine whether we can catch the bug automatically with fuzzing. The second program is included to test the behavior of switch together with a dynamic, stateful control plane. This example involves additional instrumentation in the control plane, as overviewed in \cref{sec:sut_control_instrumentation}. Both examples include components that cannot be handled by static verifiers: a hardware-only bug in the first case, and a dynamic stateful control plane in the second case.

\begin{figure}[t]
    \centering
    \subfloat[\name]{
    \includegraphics[width=\linewidth]{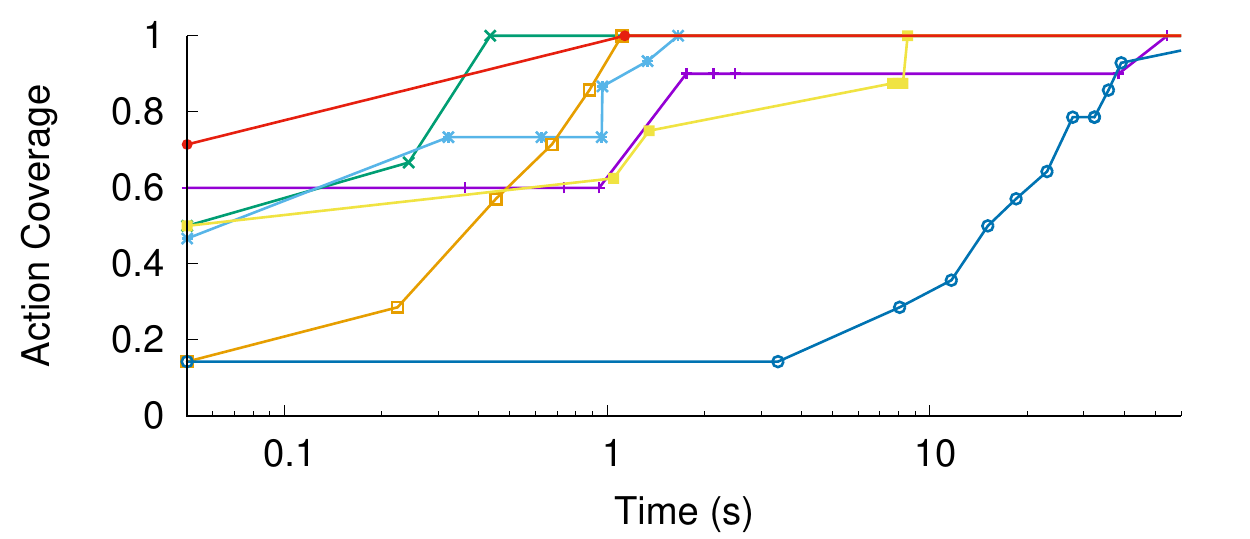}
    \label{fig:actions_vs_time_fp4}
    }
    \vspace{-1em}
    \subfloat[p4pktgen (cherry-picked table rules)]{
    \includegraphics[width=\linewidth]{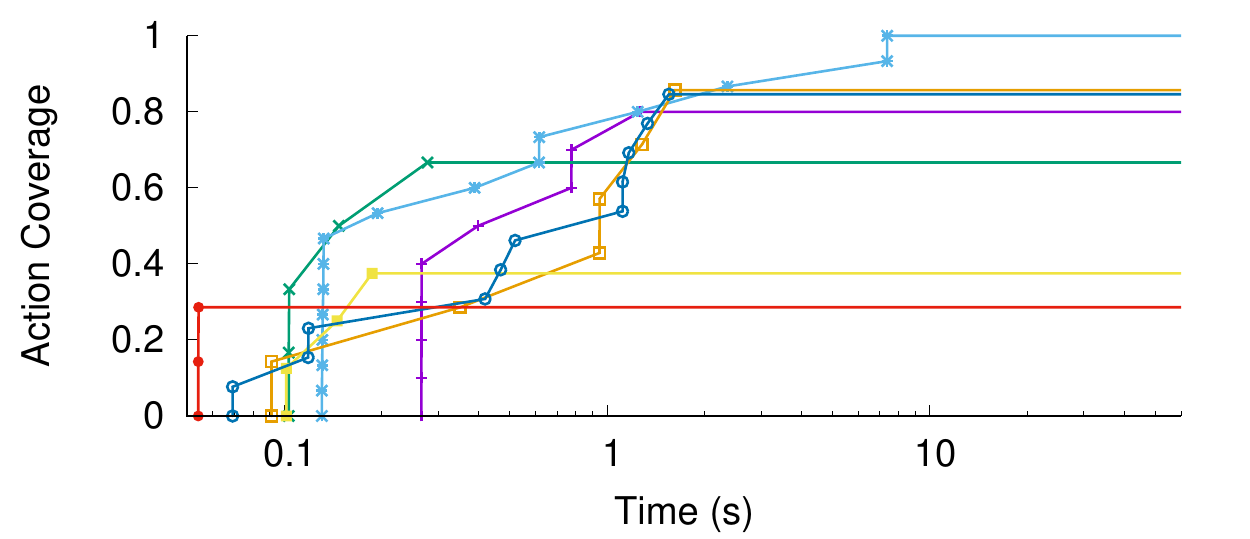}
    \label{fig:actions_vs_time_p4pktgen}
    }
    \caption{Action coverage over time for \name and p4pktgen.
    Coverage is normalized to the total number of actions in the program.
    When comparing (\subref{fig:actions_vs_time_fp4}) and (\subref{fig:actions_vs_time_p4pktgen}), we caution readers to consider the differences laid out in \cref{sec:eval_coverage}.}
    \label{fig:actions_vs_time}
\end{figure}

\begin{figure}[t]
    \centering
    \subfloat[\name]{
    \includegraphics[width=\linewidth]{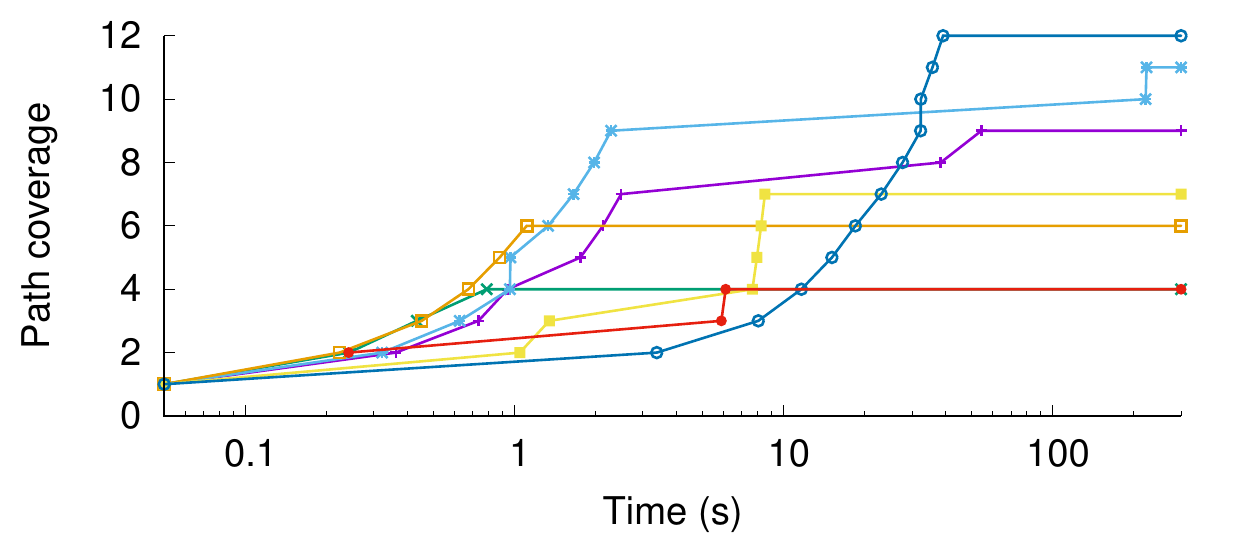}
    \label{fig:paths_vs_time_fp4}
    }
    \vspace{-1em}
    \subfloat[p4pktgen (cherry-picked table rules)]{
    \includegraphics[width=\linewidth]{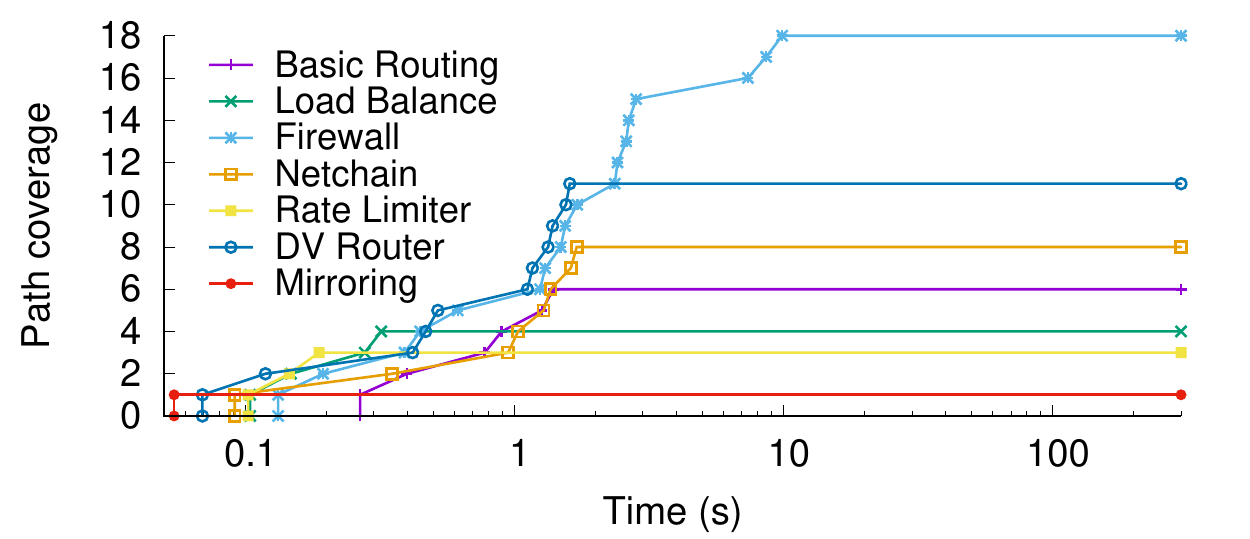}
    \label{fig:paths_vs_time_p4pktgen}
    }
    \caption{Path coverage over time for \name and p4pktgen over up to a 5\,min trace.
    When comparing (\subref{fig:paths_vs_time_fp4}) and (\subref{fig:paths_vs_time_p4pktgen}), we caution readers to consider the differences laid out in \cref{sec:eval_coverage}.}
    \label{fig:paths_vs_time}
\end{figure}

\subsection{\name Covers Code Quickly}
\label{sec:eval_coverage}

We begin by evaluating the speed at which \name can provide code coverage and test programmable switches.
We ran \name over all of the programs described in \cref{tab:test_programs} on the setup described in \cref{sec:implementation}, and we log every time a packet arrives at the SDT control plane with a newly covered path through the P4 program or state.

\cref{fig:actions_vs_time_fp4} shows the speed at which \name triggers all actions in the target programs.
The y-axis is the fraction of unique actions triggered divided by the total number of unique table-action pairs in the programs.
We note that, in all but one of our test programs, \name provides complete coverage for all the programs within around 1\,min.
This is true even for programs with only a single entry in a table with a wide keyspace.
The only exception was DV Router, where an action hit depends on (1) a properly formatted incoming routing update followed by (2) a matching ARP response for the next-hop router and (3) a packet that hits the target routing table entry.
While \name can eventually trigger this sequence of events, step (2) is currently improbable as it relies on a metadata field (\code{nextHop}) whose dependencies cross the DP/CP boundary.
\cref{fig:paths_vs_time_fp4} also shows results from the same run for paths through the program, defined as a either a unique sequence of triggered actions or a change in the control-plane state.

We also show results for another open-source P4 fuzzer, p4pktgen~\cite{p4pktgen}.
We note an important difference between the experiments: p4pktgen uses symbolic execution to solve for a small number of input packets and assumes it can freely configure the control plane rules when doing so.
As such, some of the inputs/configurations are not actually achievable in real networks with real control planes.
In fact, we needed to remove all \code{default\_action} statements from the test programs (we modified p4pktgen so that it would stop considering impossible \code{NoAction} actions).

In contrast, \name takes the deployed program and its existing table entries and configurations.
Despite this advantage, \cref{fig:actions_vs_time_p4pktgen,fig:paths_vs_time_p4pktgen} demonstrate that p4pktgen still struggles to achieve full action coverage in real programs.
While they can achieve some coverage very quickly, limitations in its current language support (e.g., statefulness and egress logic) mean that some paths are never solvable.

Note that the higher path coverage of p4pktgen in \cref{fig:paths_vs_time_p4pktgen} is due it taking impossible paths.
For example, \code{Firewall} contains a IPv4 lookup followed by per-port map lookup.
p4pktgen finds paths that include an IPv4 miss (where the packet is dropped) followed by a map hit on a non-existent port.
The static table rules do not allow \name to take this path.

\subsection{Factor Analysis}
\label{sec:eval_factors}

\begin{figure}
    \centering
    \includegraphics[width=\linewidth]{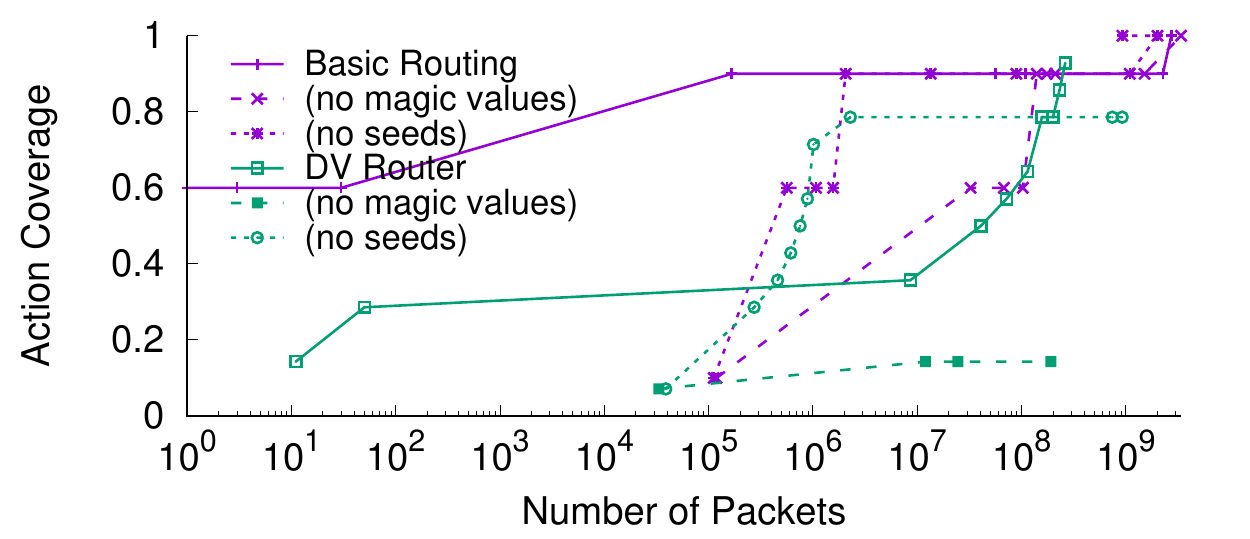}
    \caption{Packet efficiency
    (i.e., actions covered, relative to total, per test packet)
    with and without \name's optimizations.}
    \label{fig:packet_efficiency}
\end{figure}

To evaluate the benefits of different features of \name, we evaluate the packet efficiency of \name in covering the P4 program over a 5\,min period.
\cref{fig:packet_efficiency} shows these results for (1) the full \name implementation, (2) \name without magic values (the mutations of \cref{sec:sdt_mutations} that are targeted for specific table entries or conditional statements), and (3) \name without coverage-guided fuzzing (and only the parser-based seed generation).
We show values for two programs that illustrate different effects.
Basic Routing benefits from magic values to quickly cover most actions, but hitting the remaining actions relies solely on the long tail of randomness.
DV Router, on the other hand, is aided by \name's optimizations in both its initial (first 100 packets) and final coverage.

\subsection{Overhead of \name}
\label{sec:eval_overhead}

\name's primary overheads stem from the instrumentation it adds to the SUT in order to gain enough visibility to implement its greybox fuzzing approach.
We note that the resource consumption of the SDT is less important as the program is not co-resident with any other programs.
Rather, \cref{tab:resources} shows the key resources required on the SUT, which are generally low across all tested programs.
The overhead does mean that not all programs are amenable to \name's greybox approach.
An exploration of which of \name's features can be relaxed to address programs that are already close to exhausting all resources is out of the scope of this work.

\begin{figure}[!t]
\begin{mdframed}[style=background]
\begin{lstlisting}[language=P4,xleftmargin=1.5ex,xrightmargin=1.5ex]
table #tiMirror# {
    reads { ethernet.etherType : exact; }
    actions { #aiMirror#; }
    default_action: #aiMirror#(); 
}
\end{lstlisting}
\end{mdframed}
\vspace{-1em}
\caption{A snippet of a table that triggers the mirroring bug.}
\label{lst:mirroring}
\end{figure}

\subsection{Case Studies}
\label{sec:eval_case_studies}
We now present our experiences testing and finding issues in real programs using \name.

\subsubsection{Mirroring Bug}
\label{subsec:mirroring-bug}

As previously mentioned, during the development of \name, we stumbled on a bug in the interaction between mirroring and default actions.
Specifically, given a table such as \cref{lst:mirroring}, if the action \code{aiMirror()} configures the packet to be mirrored, the P4 specification would suggest that the packet should be mirrored even if the user does not add any entries beyond the default entry.
In reality, while other primitive actions in \code{aiMirror()} execute properly, we find that packet is not multicasted.
Only after adding a non-default entry does the multicast properly apply.
After adding an assertion to the input program (Mirroring), \name finds the path that violates the assertion in under 1 seconds.

While this bug was identified and addressed in more recent versions of the switch SDE, static verification tools that only consider the P4 program itself would not have found any faults in this program.
Further, we note that the simulators provided as part of the behavioral model and the switch SDE also do not catch this bug.
Only when this program is deployed to a hardware switch will this issue manifest.

\subsubsection{Testing a Distance Vector Router}

We also use \name to test for issues in a device in which the P4 data plane and its control plane interact to implement a distance-vector powered IPv4 router.
The router provides the basic functionality required for it to be placed in an arbitrary network and learn its surroundings.
Thus, the router's functionality includes the ability to handle ARP requests and responses, to understand Ethernet forwarding, and to execute a distance vector protocol to determine the correct set of LPM routing table entries.
Like a real router, the data plane is designed to be general to the topology, port counts, and address assignments of the network; instead, the control plane will configure all tables based on a provided interface configuration and data learned from neighbors or passing packets.

The original purpose of this router is as a teaching tool: students are given some skeleton code and are expected to fill in the P4 and control-plane logic for the above protocols.
Testing and debugging students' implementations is a critical task.
While testing the end-to-end correctness of the implementation is straightforward (e.g., by deploying a set of their routers and connecting two Linux hosts to either side), it is often also useful to test for violations of basic invariants, which can serve as both sanity checks and a method to localize errors.
Because of the tight integration of the control and data plane, static verification tools are not sufficient and prone to false positives.
In particular, there are many invariants where one set of table rules may result in correct behavior and another that result in errors.
For example:

\begin{vinlist}
\item Testing that the router only responds to L2 frames destined for the local interface's MAC address or the broadcast address.
\item Testing that outgoing packets are all filled with a sender MAC address corresponding to the egress port.
\item Testing that outgoing ARP responses match the incoming requests (e.g., the correct operation code and sender/target hardware and protocol addresses.
\end{vinlist}

Further, the target includes routing packets that are generated by the onboard control plane, and while the data plane \textit{could} produce errors depending on the contents of the generated packets, they \textit{will not} because of the correctness of the control plane.
Example of such properties include:

\begin{vinlist}
\item Testing that outgoing distance-vector updates include the correct source IP.
\item Testing that the outgoing distance-vector updates are properly formatted, e.g., the advertised cost is less than some preconfigured `small infinity' so as to mitigate the count-to-infinity problem of distance vector protocols. 
\end{vinlist}

\noindent
In all cases, \name enables expressive testing of the complete programmable switch that only finds bugs that are feasible.

\section{Related Work}

\topheading{Fuzzing and testing for programmable dataplanes}
Prior to \name{}, p4pktgen~\cite{p4pktgen},
P4RL~\cite{shukla2019p4rl},
and
P6~\cite{shukla2021fixp6}
were the first to propose fuzzing for P4 dataplanes.
These tools demonstrated that P4 fuzz testing could be effective at identifying a wide variety of bugs. Going off the past observation from software fuzzing that more tests equals more bugs~\cite{aflfast}, \name{} leverages programmable switches to generate test packets up to 6 orders of magnitude faster than the prior P4 fuzzing work that is based on software emulation. Moreover, emulating the P4 switch in software means that most prior works cannot detect bugs that only show up in hardware (an observation also made in~\cite{leveraging2019}) such as the mirroring bug from \S\ref{subsec:mirroring-bug}.

Work on P4 fuzzing builds on prior research in automatic test packet generation (ATPG~\cite{zeng2012atpg}), including tools such as PAZZ~\cite{shukla2019pazz} and Pronto~\cite{zhao2017pronto}.
\name{} shares a similar goal to these works (i.e., automatically identifying bugs), but specifically takes advantage of the programmability of P4 to enable efficient greybox (rather than blackbox) testing by tracking the test coverage of packets in the dataplane itself.

P4Fuzz~\cite{agape2021p4fuzz} and Gauntlet~\cite{ruffy2020gauntlet} propose fuzz testing for P4 compilers by generating structured test P4 programs more in the style of traditional compiler testing~\cite{csmith}. These works can identify a variety compiler bugs across a diverse set of P4 programs. In contrast \name{} focuses more narrowly on rigorously testing a \emph{particular} program in its entirety, including the compiler,  hardware, control plane, and data plane.

\heading{Static P4 verification}
Static verification offers an alternative approach to finding bugs in P4 programs.
There is a long line of prior work on applying static verification techniques
such as symbolic execution~\cite{dumitrescu2019netdiff,neves2018verification,stoenescu2018debugging,10.1145/3230543.3230582,lopes2016automatically} and model checking~\cite{tian2021aquila} to P4. 
Symbolic execution techniques can be categorized based on the correctness specification; for example,
netdiff~\cite{dumitrescu2019netdiff} uses differential testing~\cite{mckeeman1998differential,groce2007randomized} to define correctness, and
Assert-P4~\cite{10.1145/3185467.3185499}
uses an expressive assertion language.
Systems like bf4~\cite{10.1145/3387514.3405888} combine static verification with additional runtime checks to ensure properties that it can not verify statically.

Static verification offers strong guarantees regarding completeness---all possible input packets are checked for correctness---but it is not a panacea. Static verifiers can not yet (1) prove properties about \emph{stateful} programs that would involve reasoning about potentially arbitrarily large \emph{sequences} of packets, and they can not (2) find bugs in the switch hardware, (3) P4 compiler or (4) the control plane. Finally, because these tools lack visibility into the control plane, they may (5) report false positives~\cite{10.1145/3387514.3405888} by conservatively overapproximating the control plane's actual behavior.

We view \name{} as being complementary to static verification tools. It can test a switch program \textit{in vivo} but provides no coverage guarantees. To improve coverage in practice, \name{} leverages a combination of (1) a design based on leveraging high-throughput programmable switches, (2) P4 instrumentation for coverage guided feedback, and (3) a bevy of other optimizations (See \S\ref{sec:sut} and \S\ref{sec:sdt}).







\heading{Stateful fuzz testing}
Finally, \name{} falls into a broad category of research in bringing fuzz-testing to complex stateful systems. Popular general-purpose stateful fuzzers include RESTler~\cite{atlidakis2019restler}, Ijon~\cite{aschermann2020ijon}, Peach~\cite{Peach2016}, BeSTORM~\cite{BeStorm}, and Sulley~\cite{Sulley2014}. Outside of programmable dataplanes, stateful fuzzing has also been successfully applied to stress-test the security of network protocols~\cite{BooFuzz2022,aflnet,gascon2015pulsar,natella2021profuzzbench,banks2006snooze,abdelnur2007kif,tsankov2012secfuzz}).
%
Compared to these works, \name{} can not efficiently restart the switch, and thus opts not to generate sequences of test inputs. Instead it leverages the observation that most P4 programs are meant to be run in a ``continuous'' mode and naturally reset switch state. \name{} also collects control plane state change information and incorporates this information to derive new seed inputs.
%
\section{Limitations and Future Work}

In principle, by virtue of randomness, \name is eventually able to trigger any possible data plane bug except those that involve unused fields or that only manifest when incoming packets are rare.
One could also cover those types of bugs, e.g., if \name had support for dropping a random number of seed packets, but doing so would greatly sacrifice throughput.

Looking forward, we note that there are significant opportunities for extending \name to incorporate more recent advancements in the field of fuzzing, and/or to more efficiently cover several classes of bugs that are currently inefficient for \name to catch.
The space of possible optimizations is infinite,
but promising directions include: coordinated time stamp emulation in the data plane and control plane to better handle timing-triggered bugs, pruning of old or useless seeds to improve the efficiency of exploration, prioritization of existing seeds to target known gaps in the program coverage, gating seed packets based on the current system state to guarantee uniqueness of seed coverage, and tracking control plane code coverage to more efficiently explore bugs that originate there.

\section{Conclusion}

In this paper, we present \name, the first line-rate greybox fuzzing framework for P4 programmable switches.  
\name adapts several carefully selected, time-tested ideas from the realm of traditional application fuzz testing and demonstrates how to adapt the ideas to programmable switches---with the switches serving as both the target and the fuzzer.
Our evaluation demonstrates that \name can quickly find bugs in programs, even if the bugs are not in the P4 program itself (e.g., in the case of compiler and control-plane bugs).

{
\small
\balance
\bibliographystyle{plain}

}

\end{document}